\documentclass[twocolumn,twocolappendix]{aastex63}  

\usepackage{amssymb,amsmath,bm}
\usepackage{graphicx}
\usepackage{natbib}
\usepackage{color}
\usepackage{soul}

\usepackage{algorithm}
\usepackage[noend]{algpseudocode}
\usepackage{tikz}
\usetikzlibrary{quantikz}
\usepackage{savesym}
\savesymbol{splitbox}
\usepackage{adjustbox}
\restoresymbol{TXF}{splitbox}
\usepackage{hyperref}

\def\aap{A\&A}

\def\apjs{ApJS}
\def\apjl{ApJL}

\received{xxx xx, 2025}
\revised{xxx xx, 2025}
\accepted{xxx xx, 2025}
\submitjournal{ApJ}

\begin{document}

%
%
%
%


\title{SPINN: Advancing Cosmological Simulations of Fuzzy Dark Matter with Physics Informed Neural Networks}

\correspondingauthor{Ashutosh K. Mishra}
\email{ashutosh.mishra@epfl.ch}

\author[0009-0002-8819-8236]{Ashutosh K. Mishra}
\affiliation{Institute of Physics, Laboratory of Astrophysics, École Polytechnique Fédérale de Lausanne (EPFL), Observatoire de Sauverny, Versoix, 1290, Switzerland}
\author[0000-0002-1027-1213]{Emma Tolley}
\affiliation{Institute of Physics, Laboratory of Astrophysics, École Polytechnique Fédérale de Lausanne (EPFL), Observatoire de Sauverny, Versoix, 1290, Switzerland}


\begin{abstract}
Physics-Informed Neural Networks (PINNs) have emerged as a powerful tool for solving differential equations by integrating physical laws into the learning process. This work leverages PINNs to simulate gravitational collapse, a critical phenomenon in astrophysics and cosmology. We introduce the Schr\"odinger-Poisson informed neural network (SPINN) which solve nonlinear Schrödinger-Poisson (SP) equations to simulate the gravitational collapse of Fuzzy Dark Matter (FDM) in both 1D and 3D settings. Results demonstrate accurate predictions of key metrics such as mass conservation, density profiles, and structure suppression, validating against known analytical or numerical benchmarks. This work highlights the potential of PINNs for efficient, possibly scalable modeling of FDM and other astrophysical systems, overcoming the challenges faced by traditional numerical solvers due to the non-linearity of the involved equations and the necessity to resolve multi-scale phenomena especially resolving the fine wave features of FDM on cosmological scales.

\end{abstract}
\keywords{gravitation --- dark matter --- methods:numerical}
\section{Introduction}\label{intro}
Studies of the Cosmic Microwave Background (CMB)
suggest that the baryonic matter content of the universe constitutes only about 5\% of the total energy-matter density \citep{2020A&A_LCDM}. The remaining invisible components include ~68\% dark energy, driving the accelerated expansion of the observable universe \citep{1998_Riess}, and ~27\% a non-luminous, non-interacting, non-baryonic matter called \emph{dark matter}. Dark matter plays a crucial role in the well-accepted scenario of hierarchical cosmic structure formation \citep{1978_White,2003_Zhao}. 
In particular, it drives the evolution of density fluctuations, as dark matter is the dominant matter component in the Universe. The primordial density fluctuations (on the order of 
$10^{-5}$) observed in the CMB seed the gravitational growth of dark matter fluctuations through hierarchical clustering, followed by the infall of baryons into the resulting potential wells after recombination, leading to structure formation.
Dark matter also holds galaxies together within clusters \citep{1980_Rubin,2006_Clowe}. 
Therefore, studying the nature of dark matter (DM) is essential to understand the formation and evolution of cosmic structures.

In most theoretical and numerical treatments DM is treated as a cold collisionless fluid called Cold DM (CDM). CDM has been indispensable in explaining various cosmological observations spanning a wide range of scales \cite[e.g.][]{2012_FandW,2012MNRAS.425..415S,2012MNRAS.427.3435A}. On small cosmological scales (ranging from tens of parsecs to a few kiloparsecs), however, its validity is questionable given various identified problems like the \lq{core-cusp problem\rq}, the \lq{too-big-to-fail problem}, and so on \cite[e.g.][]{2015_Weinberg,2017_B, 2022_Sales}. This may be attributed to the complex baryonic feedback mechanisms, and/or the unknown particle nature of CDM. Furthermore, certain outliers, such as those observed in dwarf galaxy rotation curves, remain unexplained even when baryonic effects are incorporated into simulations \citep{2015_Oman, 2022_Sales}.

In addition to these cosmological discrepancies, no evidence has been found for popular CDM candidates in terrestrial particle physics experiments \citep{2024_Direct_det}, in particular weakly interacting massive particles (WIMPs). Furthermore, very strong limits on the abundance of MACHOs (Massive Compact Halo Objects) have been placed by the Event Horizon Telescope (EHT) observations \citep{PhysRevD.105.063506}. As a result, alternative dark matter models based on ultralight (\lq{axion-like\rq}) scalar particles have come under the spotlight. The ultralight particles have masses typically around m $\sim 10^{-20} - 10^{-22}$ eV, implying the presence of quantum effects causing wave-like behaviour at small (i.e. galactic or kpc) scales. These effects are in turn claimed to help solve small-scale problems of $\Lambda$CDM; 
For example, ultralight DM should form solitonic cores at the center of dark matter haloes  \cite[e.g.][]{2014_Schive,2017_Hui}, with a predicted density profile that matches common dwarf-spheroidal galaxies and our own Milky Way galaxy  \citep{DEMARTINO2020100503}. In addition, these ultralight scalar particles are generic predictions of various particle physics theories ranging from quantum chromodynamics (QCD) \citep{P_Q} to  String theories \cite[][]{2010_Axiverse,2015_Marsh,2021_Skivie}. 

DM composed of ultralight scalars is often called fuzzy dark matter (FDM) for its quantum wave phenomenology. FDM is particularly interesting as it has astrophysically-relevant de Broglie wavelength spanning from tens of parsecs to a few kiloparsecs with rich phenomenological implications \citep[See for a review,][]{2021_Hui}. In addition, they retain the CDM-like behaviour at large scales while suppressing structure formation on small scales. This makes the FDM model a potential candidate for addressing small-scale challenges of the $\Lambda$CDM model. In the non-relativistic regime, the dynamics of FDM is governed by the SP equations, though relativistic effects can also be considered in more detailed models.

In any DM model, studying the evolution of DM density perturbations is crucial for understanding structure formation in the universe. The most commonly used approach to investigate the gravitational instability of these density perturbations in a collisionless fluid is through the N-body discretization of the fluid's phase-space distribution function, as described by the Vlasov-Poisson (VP) system of equations, which governs CDM dynamics. A recent study by \citet{2018_Mocz}, demonstrates the numerical correspondence between the 6D+1 Vlasov-Poisson equations and the 3D+1 Schrödinger-Poisson (SP) equations in cosmological simulations, further motivating the study of DM models that can be described by the SP equations.


Historically, studies of FDM dynamics relied on linear theory \citep[][]{Wayne_2000} or warm dark matter simulations incorporating an appropriate transfer function \cite[e.g.][]{2022_paduroiu,2025_Liu}. However, nonlinear effects play a crucial role, significantly impacting the power spectrum \citep[]{2021_May}, necessitating the use of full numerical simulations to make further progress in understanding FDM dynamics. Numerical studies have analyzed structure formation in FDM model, utilizing both the wave formulation (SP Equations) and fluid (or Madelung) formulation. They have employed a wide range of algorithms including but not limited to spectral methods \cite[e.g.][]{Mocz_2017_BECDM,2018_Du}, finite difference methods \cite[See,][]{2014_Schive,2020_Mina}, and so on. 

A primary limitation in wave-based simulations of FDM is the necessity to resolve the de Broglie wavelength, even in regions where the density field is smooth and non-vanishing. For a typical FDM particle of mass m $\sim$ $10^{-22}$ eV and a velocity v = 100 km/s, the de Broglie wavelength ($\sim$1.2 kpc) is significantly smaller than the box size required for a cosmological simulation of relevant size, making it very difficult to compare to conventional CDM simulations which require high resolutions only in dense regions. As a result, FDM simulations have been constrained to relatively small box sizes of few Mpc. \citet{2021_May} achieved a box size of 10 Mpc/h for a SP simulation of FDM with m = $7 \times 10^{-23}$ eV, though they could only evolve the simulation to z=3 due to insufficient resolution.
On the other hand, simulations employing the fluid formulation do not need to resolve de Broglie wavelength to obtain large scale dynamics; instead, they incorporate a quantum pressure term using smoothed particle hydrodynamics methods. As a result, these simulations are comparable to CDM  N-body simulation results \cite[][]{2016_Schive,2018_Nori} in terms of resolution and box size. But this method has its own challenges due to vanishing densities leading to ill-defined quantum pressure terms making it unable to capture quantum interference effects \cite[][]{2018_Zhang,2019_Li}. Regardless of the formalism, the time step in FDM simulations scales quadratically with the grid size ($\Delta t \propto \Delta x^2$), making high-resolution simulations computationally expensive and time-consuming. Recently, \citet{2024_Kunkel} have managed to develop a hybrid numerical scheme with both these formulations allowing for more accurate zoom-in simulations for cosmological volumes, than any single formulation based approaches. The box size achieved though is still about 5.6 Mpc/h for hybrid cosmological simulation and 10 Mpc/h for power spectrum tests. However, these sizes are not directly comparable to the much larger CDM simulation box sizes, which range from hundreds to thousands of Mpc/h.

With limitations of current numerical schemes to simulate FDM dynamics to equivalent volumes of CDM simulations, the need for accelerated large box FDM simulations has become important to constrain the micro-physical nature of dark matter. Deep learning has been leveraged to accelerate cosmological simulations, and convolutional neural networks (CNNs) have been successfully applied to correct simple numerical approximations~\citep{doi:10.1073/pnas.1821458116,dai2020,necola}, predict radiative transfer and reionization maps~\citep{chardin_deep_2019,pinion}, or paint galaxy stellar masses on top of DM fields~\citep{dm2gal, baryonicifcation}. However, the usefulness of deep neural networks is limited by  available data. Neural networks require training on datasets with ample statistical representation, typically containing thousands of instances of relevant features or classes. However, even when trained on large datasets, these models may not inherently adhere to the underlying physics of the problem, leading to potential extrapolation errors, generalization issues, and unreliable predictions \citep{whytrust}. Furthermore, the inherent opacity of deep neural network inference makes it challenging to interpret and fully trust their outputs.


Physics informed machine learning can address these shortcomings and improve the reliability and generalizability of deep learning approaches by enforcing physical laws such as symmetry, conservation, or dynamics defined by time-dependent and non-linear partial differential equations. In particular Physics Informed Neural Networks (PINNs) have been a powerful alternative for solving differential equations in both forward and inverse problems \cite[e.g.][]{2021_Moya,2022_Jagtap,2024_Baty,2024_Matt,2025_Wu}.PINNs have been sucessfully leveraged to accelerate numerical simulations in a variety of domains. They have been used to solve the Navier-Stokes equations and reproduce 3D fluid flow simulations with incredible accuracy~\citep{HFM, JIN2021109951}, the PINN Euler equations (EE) that model high-speed aerodynamic flows~\cite{pinneuler}, heat transfer~\citep{pinnheat}, and radiative transfer~\citep{MISHRA2021107705}.
Even though, PINNs have been extensively and successfully used for fluid, plasma, and climate simulations \citep{2020_Mao,f-PICNN,2023_fusion_plasma}, they have been mostly unexplored for cosmological simulations mainly due to the difficulty in modelling long-range effects of gravitational potential.

In this work, we explore for the first time the potential of PINNs for cosmological simulations, in particular to achieve efficient simulations of FDM. 
We begin by simulating structure collapse of FDM in 1 dimension and then extend to 3 dimensions, simultaneously getting insights from the 1D results. We believe PINNs' success in modelling this gravitational collapse is an incentive for its use to create a full-fledged cosmological FDM simulation using PINNs.

This paper is structured as follows. In Section \ref{Bkgd}, we present the governing equations for the Fuzzy Dark Matter used in this work and the idea of PINNs to provide the necessary background. In Section \ref{method}, we outline the methods, including the use of PINNs for SP equations and the corresponding architectures, explaining their selection and relevance. We present the results and lead the discussion in Section \ref{Res} before we conclude in Section \ref{conc}. The details of numerical benchmark (\ref{appendix_A}), comparison with fluid formalism (\ref{appendix_B}), and additional tests results (\ref{appendix_c})  are left in the appendix.

\section{Theoretical Background}\label{Bkgd}
\subsection{Governing Equations}\label{GE}
 Wave DM  in simplest terms is an ultralight particle with  mass $m \ll 30$ eV. Given the measured local DM  mass density of $\sim 0.4$ GeV cm$^{-3}$, such particles must necessarily be bosonic. A scalar-field boson is described by its complex-valued wavefunction
 $\Psi(\boldsymbol{x},t)  \in \mathbb{C}$ with $\boldsymbol{x}\in \mathbb{R}^{3}$ such that $|\Psi(\boldsymbol{x},t)|^2 = \rho/\bar{\rho}$ where the background density $\bar{\rho}$ is defined considering normalization of the wavefunction over the domain volume $\mathcal{V}$ (for a cubic box of length L in 3D, it would be $L^3$),
\begin{align}
    \frac{1}{\mathcal{V}} \int d\mathcal{V}  |\Psi|^2 = 1 .
\end{align}
This wavefunction obeys the Schr\"odinger equation,
\begin{equation}
    \label{SE}
    i\hbar \partial_t \Psi = -\frac{\hbar^2}{2m} \nabla^2 \Psi + mU \Psi 
\end{equation}
where $m$ is the boson mass, $\hbar$ is the reduced Planck constant, and $U$ is the gravitational potential defined by the Poisson equation,
\begin{equation}
\label{PE}
    \nabla^2 U = 4\pi Gm(\rho - \bar{\rho})
\end{equation}
 where $G$ is the gravitational constant and $\bar{\rho}$ is the background density (in our case the average density over the considered volume). Equations (\ref{SE}) and (\ref{PE}) comprise the SP system of equations that governs the evolution of FDM \citep{2021_May}.
 

 It should be noted that the Eq.(\ref{SE}) describes the evolution of a single macroscopic wavefunction of a Bose-Einstein condensate, with a mass density $|\Psi|^2 = \rho/\bar{\rho}$,   rather than the wavefunction of an individual particle. Identifying $\lambda = \hbar/m$ as the intrinsic scale of the problem which also determines the onset of quantum effects in the system, we can rewrite the above system of equations as:
\begin{align}\label{SE_re}
i \frac{\partial}{\partial t}\Psi(\boldsymbol{x},t) &= \bigg(-\frac{\lambda}{2} \nabla^2 + \frac{1}{\lambda} V\bigg)\Psi(\boldsymbol{x},t)
\\
\label{PE_re}
 \nabla^2V(\boldsymbol{x},t) &= (|\Psi(\boldsymbol{x},t)|^2 - 1) 
\end{align}
Where, $V = U/\alpha$ is the dimensionless potential with $\alpha = 4\pi G \bar{\rho}$.

In this paper we choose the relevant constants ($G$, $\bar{\rho}$, $L$, where $L$ is the computational domain size) such that $\alpha = 1$. The characteristic length $\lambda$ determines the strength of the potential term in the Schr\"odinger equation. As $\lambda \to \infty $ the gravitational potential term vanishes leaving us with the free Schr\"odinger-equation which just spatially smoothens out the initial density distribution. On the other hand, as $\lambda \to 0$ the potential term dominates and the evolution becomes highly non-linear through its coupling to Poisson-equation. In this case it leads to the collapse of the structure under the influence of the gravitational potential.

Alternatively, the SP equations can be re-written using the \citet{1927_Madelung} transformation:
\begin{align}
    \Psi = \sqrt\rho e^{iS}
\end{align}
where $\rho$ and $S$ are the density and phase of the wavefunction. We can rewrite the SP equations in terms of $\rho$ and $S$ to obtain the Hamilton-Jacobi-Madelung equations:
\begin{align}
    \frac{1}{\lambda}\partial_t \rho + \nabla \cdot (\rho \nabla S) = 0 \\
    \frac{1}{\lambda}\partial_t S + \frac{1}{2} (\nabla S)^2 + \frac{1}{\lambda^2}V -\frac{1}{2}\frac{\nabla^2 \sqrt\rho}{\sqrt\rho} = 0\\
    \nabla^2 V - (\rho-1) = 0
\end{align}
Identifying the velocity as $\mathbf{v} = \lambda \nabla S$ we obtain the Madelung equations for pressureless, compressible fluid flow in a gravitational potential modified by the quantum pressure term $\nabla^2 \sqrt\rho/\sqrt\rho$:
\begin{align}
    \partial_t \rho + \nabla \cdot (\rho \mathbf{v} ) = 0 \\
    \partial_t \mathbf{v} + \mathbf{v}\cdot \nabla\mathbf{v}  + \nabla V -\frac{1}{2}\lambda^2\nabla\frac{\nabla^2 \sqrt\rho}{\sqrt\rho} = 0\\
    \nabla^2 V - (\rho-1) = 0
\end{align}
The equations, similar to the Vlasov-Poisson equations governing CDM in the limit 
$\lambda \to 0$ and before shell crossing, are recovered once the quantum pressure term vanishes which happens on large scales \citep{2002_Bernardeau,2021_Hui}. Despite the fact that these equations are highly non-linear, they become tractable with existing hydrodynamic codes. However, it is inherently ill-behaved due to the presence of quantum pressure term which blows up as $\rho \to 0$, which is common in the DM overdensity evolution where the density vanishes, especially in the voids between halos and filaments. Therefore we choose to evolve the SP-system of equations (\ref{SE_re})-(\ref{PE_re}) instead of the Madelung formalism in our work.

\subsection{Physics Informed Neural Networks(PINNs)}
Physics Informed Neural Networks (PINNs) are neural networks for solving differential equations that incorporate physical laws into the learning process \citep{2019_Raissi}. They are often used to simulate dynamical systems where the governing equations are PDEs. The standard procedure is to train the neural network NN($X,\theta$) as a surrogate model which approximates the PDE solution by minimizing the residual loss obtained from satisfying the PDE constraints together with the boundary and initial conditions. NN($X,\theta$) is defined by the set of parameters $\theta$ (the weights and biases of the network), over the d+1 dimensional computational space-time domain (d spatial + 1 temporal dimension) in the form of collocation points $X:[\mathbf{x},t])$. These points in turn are provided as inputs to the PINN which then learns the solution by minimizing the loss function.

The major distinction of PINNs from other ML models is in the inclusion of physics in the loss function itself. This is a form of unsupervised learning, as aside from the initial and boundary conditions no real data or analytical solution is exploited to influence the training of the PINNs. Traditional supervised learning can be implemented with the typical data loss term which incorporates the deviation of the network prediction from the real data (or in cases the available analytical solutions). In this work, we mostly use unsupervised PINNs making use of physical equations as well as the boundary and initial conditions without any inclusion of labeled data (or any additional observational features or simulations), and briefly explore semi-supervised PINNs (inclusion of simulations as additional constraint in the loss).

 The general form for the differential equation which can be solved with PINNs can be written as
\begin{align}
    \mathcal{D}[NN(X,\theta);\Lambda] = f(X), \quad X\in \Omega \\
    \mathcal{B}[NN(X,\theta);] = g(X) \quad X\in \partial\Omega
\end{align}
 where $\mathcal{D}$ is the non-linear differential operator, $\Lambda$ represents the parameters involved in the physics equations or the differential operator, $f(X)$ is the term sourcing the equations, $\mathcal{B}$ represents the operator indicating arbitrary boundary or initial conditions related to the system in consideration, $g(X)$ is the boundary function, and $\Omega \subset \mathbb{R}^{d+1} $ is the computational domain with its boundary, $\partial\Omega$. As an example, the Schr\"odinger-Poisson equations (\ref{SE_re}) \& (\ref{PE_re}) with periodic boundary conditions at time t read:
 \begin{align}
\bigg[i \frac{\partial}{\partial t} +\frac{\lambda}{2} \nabla^2 - \frac{1}{\lambda} V\bigg]\Psi(\boldsymbol{x},t) = 0 = f_1(X)
\\
 \nabla^2V(\boldsymbol{x},t) - (|\Psi(\boldsymbol{x},t)|^2)= - 1 = f_2(X) \\
 u(x_L,t)-u(x_R,t) = 0 = g_1(X)\\
 \nabla u(x_L,t)-\nabla u(x_R,t) = 0 = g_2(X)
\end{align}
where, $u := \{\Psi, V\}$ represents any of the quantities predicted using the PINN, $x_L$ \& $x_R$ are the left and right ends of the spatial dimension respectively. Boundary conditions can be Dirichlet, Neumann, Robin, or Periodic. We  restrict ourselves to periodic boundary conditions in this work to mimic real cosmological simulations.
\section{Methods}\label{method}
\subsection{SPINN}
\begin{figure*}
    \centering
    \includegraphics[width=\linewidth]{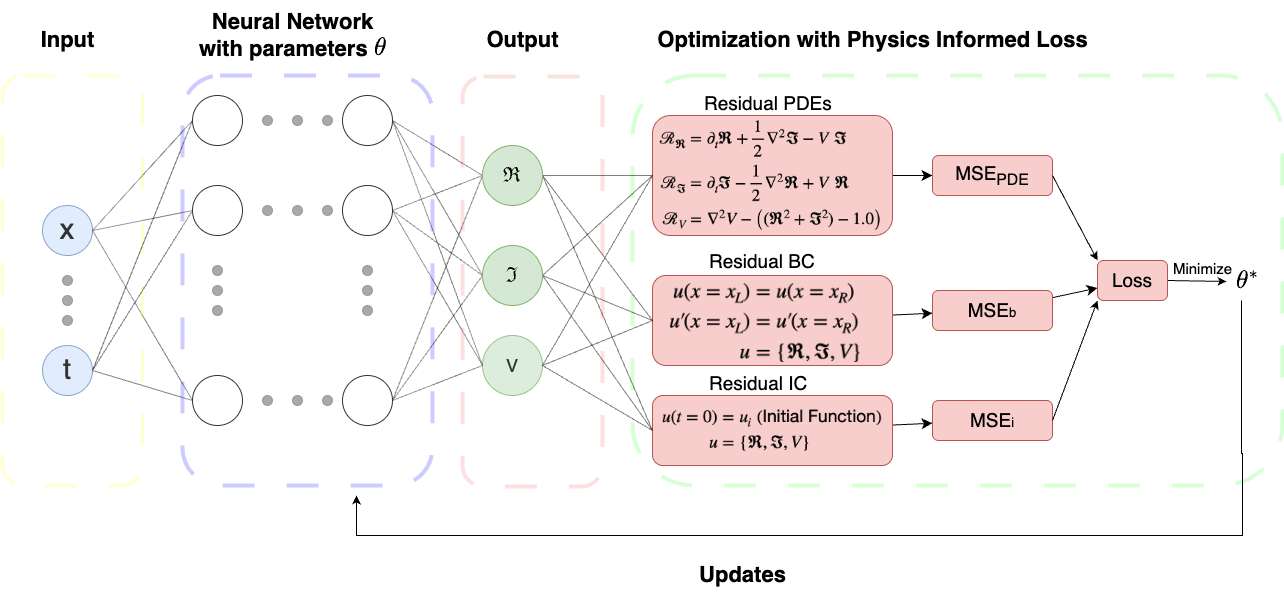}
    \caption{Schematic of the PINN-based SPINN framework: The neural network outputs $\{\mathfrak{R},\mathfrak{I}, V$\}, which serve as approximate solutions ($\text{Re}(\Psi) = \mathfrak{R},\text{Im}(\Psi)=\mathfrak{I}$). These, along with the governing PDEs, ICs, and BCs, define the total loss function. During training, the parameters $\theta$ are iteratively optimized to obtain $\theta^{*}$, yielding the final solutions of the system.}
    \label{fig1}
\end{figure*}
Here we introduce SPINN, a PINN-based model to simulate the Schr\"odinger-Poisson system in the presence of self-gravity, the dynamics governing FDM. We set $\lambda = 1$ to simplify expressions, effectively working in units where
$\hbar = m$. Writing out the real and imaginary components of the wavefunction $\Psi$ explicitly as $\Psi = \text{Re}(\Psi) + i \text{Im}(\Psi) = \mathfrak{R} + i \mathfrak{I}$, we can rewrite Eqs. (\ref{SE_re}-\ref{PE_re}) as:
\begin{align}
 \partial_t ~\mathfrak{R} &= - \frac{1}{2} \nabla^2 \mathfrak{I} + V ~ \mathfrak{I}\\
\partial_t ~\mathfrak{I} &= \frac{1}{2}\nabla^2 \mathfrak{R} - V ~\mathfrak{R}\\
 \nabla^2 V &= \mathfrak{R}^2  + \mathfrak{I}^2-1
\end{align}
As discussed in Section~\ref{GE}, our choice to train the PINNs using the Schrödinger-Poisson (SP) equations is theoretically motivated by their ability to capture the wave-like behavior inherent to FDM dynamics. In addition to this theoretical basis, we conducted empirical tests to evaluate whether a PINN architecture trained instead to predict density 
$\rho$, velocity $v$, and gravitational potential $V$ could accurately model the system's evolution. As detailed in Appendix~\ref{appendix_B}, this approach fails to reproduce key wave features of the evolution, consistent with our theoretical expectations.

Consequently, we adopt the SP formulation for training our PINNs. Specifically, we design the network to predict the real $\mathfrak{R}$ and imaginary $\mathfrak{I}$ components of the wavefunction $\Psi$, as well as the gravitational potential $V$, which is treated as an auxiliary output rather than computed through direct numerical solution. This strategy significantly reduces computational cost by circumventing the need to solve Poisson’s equation at every training step, as elaborated in Section~\ref{Res} and Appendix~\ref{appendix_c}.

We define a neural network $NN(X;\theta)$ which approximates $\mathfrak{R}$, $\mathfrak{I}$, and $V$ as follows:
\begin{align}
    NN(X;\theta) = \mathfrak{R}_{\theta}(X),\mathfrak{I}_{\theta}(X), V_{\theta}(X)
\end{align}
where $(\cdot)_{\theta}$ indicates a PINN approximation realized with a network represented by the set of parameters $\theta$. We choose to predict real $\mathfrak{R}$ and imaginary $\mathfrak{I}$ parts of $\Psi$ directly 
because it implicitly imposes the complex conjugate of the Schr\"odinger equation. This approach ensures mass conservation and correct evolution, eliminating the need to explicitly implement a continuity equation. A schematic of the same has been shown in Fig. (\ref{fig1}).

The total loss is defined as the sum of the squares of PDE residual, the boundary and the initial condition residuals at the collocation points X for each of the output variables. The PDE's residuals in terms of the output variables associated with Eqs. (\ref{SE_re}-\ref{PE_re}) for a (d+1) dimensional system are:

\begin{align}
\label{r1}    \mathcal{R}_{\mathfrak{R}}(X) &= \partial_t\mathfrak{R_{\theta}} + \frac{1}{2} \left(\sum_{i = 1}^{d} \partial_{x_i}^2 \mathfrak{I_{\theta}}\right) - V_{\theta} ~ \mathfrak{I_{\theta}}\\
\label{r2}    \mathcal{R}_{\mathfrak{I}}(X) &= \partial_t\mathfrak{I_{\theta}} - \frac{1}{2} \left(\sum_{i = 1}^{d} \partial_{x_i}^2 \mathfrak{R_{\theta}}\right) + V_{\theta} 
~ \mathfrak{R_{\theta}}\\
\label{r3}    \mathcal{R}_V(X) &= \sum_{i = 1}^{d} \partial_{x_i}^2 V_{\theta} - \left((\mathfrak{R_{\theta}}^2  + \mathfrak{I_{\theta}}^2)-1.0\right)
\end{align}
The Mean Squared Error (MSE) associated to the residuals computed in Eqs.(\ref{r1}-\ref{r3}) for a set of randomly generated collocation points sampled from a uniform distribution with each spatial dimension in the range [0,L] and temporal dimension in the range [0,$\pi$] in 3+1 dimensions $X_n^r = [x_n^r,y_n^r,z_n^r,t_n^r]$, are:
\begin{align}
    MSE_{PDE}(\theta)= \frac{1}{N_r}\sum_{n = 1}^{N_r}\Bigg[ \Big| \mathcal{R}_{\mathfrak{R}}(X_n^r) \Big|^2 +  \Big| \mathcal{R}_{\mathfrak{I}}(X_n^r) \Big|^2 + \notag \\ \Big| \mathcal{R}_V(X_n^r) \Big|^2 \Bigg]
\end{align}
Where $N_r$ is the number of collocation points chosen to evaluate the PDE residuals. The MSE associated to the residuals for boundary and initial condition for each of the outputs at randomly generated points at $X_n^b$ and $X_n^i$ are:
\begin{align}
    MSE_{b}(\theta)= \frac{1}{N_b}\sum_{n = 1}^{N_b}\Bigg[ \Big| \mathfrak{R}_{\theta}(X_n^b)-\mathfrak{R}_{b}(X_n^b) \Big|^2 +
     \notag \\
    \Big|\mathfrak{I}_{\theta}(X_n^b)-\mathfrak{I}_{b}(X_n^b)  \Big|^2 +  \Big| V_{\theta}(X_n^b)-V_{b}(X_n^b)  \Bigg]
\end{align}
\begin{align}
    MSE_{i}(\theta)= \frac{1}{N_i}\sum_{n = 1}^{N_i}\Bigg[ \Big| \mathfrak{R}_{\theta}(X_n^i)-\mathfrak{R}_{i}(X_n^i) \Big|^2 +
     \notag \\
    \Big|\mathfrak{I}_{\theta}(X_n^i)-\mathfrak{I}_{i}(X_n^i)  \Big|^2 +  \Big| V_{\theta}(X_n^i)-V_{i}(X_n^i)  \Bigg]
\end{align}
Where $N_b$ and $N_i$ are the number of collocation points on the boundary and at the initial time, respectively. The solution is then realized by minimizing the total loss ($MSE_{PDE}+MSE_b+MSE_i$) through the optimization of the neural network defined by $\theta$:
\begin{align}
    \theta^* = \underset{\theta}{\text{arg min}} \left(MSE_{PDE}+MSE_b + MSE_i\right)
\end{align}
We do not include any explicit constraints to enforce the normalization of the wavefunction. Instead, the probability conservation rather follows automatically as the correct physics is learned by the network.

As the network is trained its predictions converge to the true solution. The optimization procedure is done in a similar fashion like any other neural network with the aid of auto-differentiation ~\citep{Auto_diff} and back-propagation for gradient updates ~\citep{DL_Lecun}. The largest difference from conventional deep learning training is in the employment of mostly smooth activation functions (like sine, tanh, etc) to ensure continuous first and second derivatives of the network predictions.

In traditional numerical differentiation methods, one is constrained by the mesh size and the related truncation error. In this PINN formalism, once the training is complete, the mesh is no longer required rendering the PINN method absolutely mesh-free thanks to the use of auto-differentiation.

\subsection{Architecture and Optimization}
We begin analyzing the problem in 1D with a SPINN consisting of 2 hidden layers, each of 32 neurons as shown in Table \ref{tab:my_label1}. The network weights $\theta$ are initialized with Xavier-uniform distribution \citep{pmlr-v9-glorot10a}.
Considering the smoothness requirement for computing higher-order derivatives, we employ the \textit{sine} activation function throughout, following the formulation in \citet{2020_Siren}. Compared to more commonly used activations like \textit{tanh} and \textit{SiLU}, the periodic nature of \textit{sine} enables better representation of complex, oscillatory structures and preserves gradient quality across layers. Empirically, we observed that both \textit{tanh} and \textit{SiLU} led to reduced accuracy as compared to \textit{sine} (See appendix \ref{act_app} and Fig. (\ref{Fig2})).
Standard gradient-based optimizers like Adam \citep{2014_Kingma} are not ideal for minimizing the PDE loss due to the irregular loss landscape created by differential operators  \citep{2024_Rathore}. Therefore we adopt as advised in the same paper, a combination of Adam + LBFGS \citep{L_BFGS}, as an optimizer. We first pre-train the network with Adam for initial 200 epochs, and then switch to L-BFGS until it finally converges.  Adam is used with a learning rate of 1e-3 and L-BFGS with its default setting (a learning rate of 1 and maximum iteration of 20) along with strong-wolf line search \cite[e.g.,][]{1969_Wolfe,1971_Wolfe,Nocedal2006}. As far as the collocation points are concerned, $N_r  = N_b = N_i = 10000$ points are randomly generated from a uniform distribution in the respective space-time domain already mentioned in above Section. The hyperparameters (such as weight initialization, number of layers, number of neurons, etc) are chosen through trial and error, guided by the model’s performance in terms of minimizing the total loss, and achieving stable convergence after 1000 training iterations.
\begin{table}
    \begin{center}
    \begin{tabular}{|c|c|}
 \hline
 \textbf{Layer type} & \textbf{Activation function} \\
 \hline
 Layer 1: Dense, 32 neurons  & Sin     \\
 Layer 2: Dense, 32 neurons  & Sin    \\
 Layer 3: Dense, 03 neurons  & Linear  \\
 \hline
 \textbf{Optimizer} &  Adam + L-BFGS   \\
                     & (Adam: lr = 0.003;\\
                     & L-BFGS: with strong-wolf \\&line search)\\
\hline 
 \textbf{Loss function} & Mean Squared Error\\
 \hline
    \end{tabular}
    
    \end{center}
    \caption{General Architecture of the PINNs used in 1D case. For the meaning of each term in the table refer \citet{nielsen2015neural}. (lr := learning rate)}
    \label{tab:my_label1}
\end{table}

In three-dimensional problems, a simple neural network architecture is often insufficient. Furthermore, the selection of an appropriate activation function requires experimentation, depending on the complexity of the chosen initial function as well as the derivatives present in the PDEs . In this study, we employ a SPINN  with 5 hidden layers, each containing 16 neurons, as detailed in Table \ref{tab:my_label2}. The network weights are initialized using the Glorot-Normal (truncated Gaussian) distribution \citep{pmlr-v9-glorot10a}. The activation function for the first layer is set to the sine function, facilitating a transition into the sinusoidal input space \citep{2021_Wong}.  Subsequent layers use the \textit{Wavelet} activation function \citep{2023_Pinnsf}. This hybrid activation strategy is particularly effective given the periodic boundary conditions and initial conditions, which also involve sinusoidal components. The optimization procedure, along with the selection of collocation points, remains consistent with the approach used in the one-dimensional case.
\begin{table}
    \begin{center}
    \begin{tabular}{|c|c|}
 \hline
 \textbf{Layer type} & \textbf{Activation function} \\
 \hline
 Dense, 16 neurons  & Sin     \\
 Dense, 16 neurons  & Wavelet    \\
  Dense, 16 neurons  & Wavelet     \\
 Dense, 16 neurons  & Wavelet   \\
  Dense, 16 neurons  & Wavelet  \\
   Dense, 3 neurons  & Linear  \\
\hline
    \end{tabular}
    \end{center}
    \caption{General Architecture of the PINNs used in 3D case. For the meaning of each term in the table refer \citet{nielsen2015neural} and the activation Wavelet is described in \citet{2023_Pinnsf}.}
    \label{tab:my_label2}
\end{table}

\begin{figure*}
    \centering
    \includegraphics[width = \linewidth]{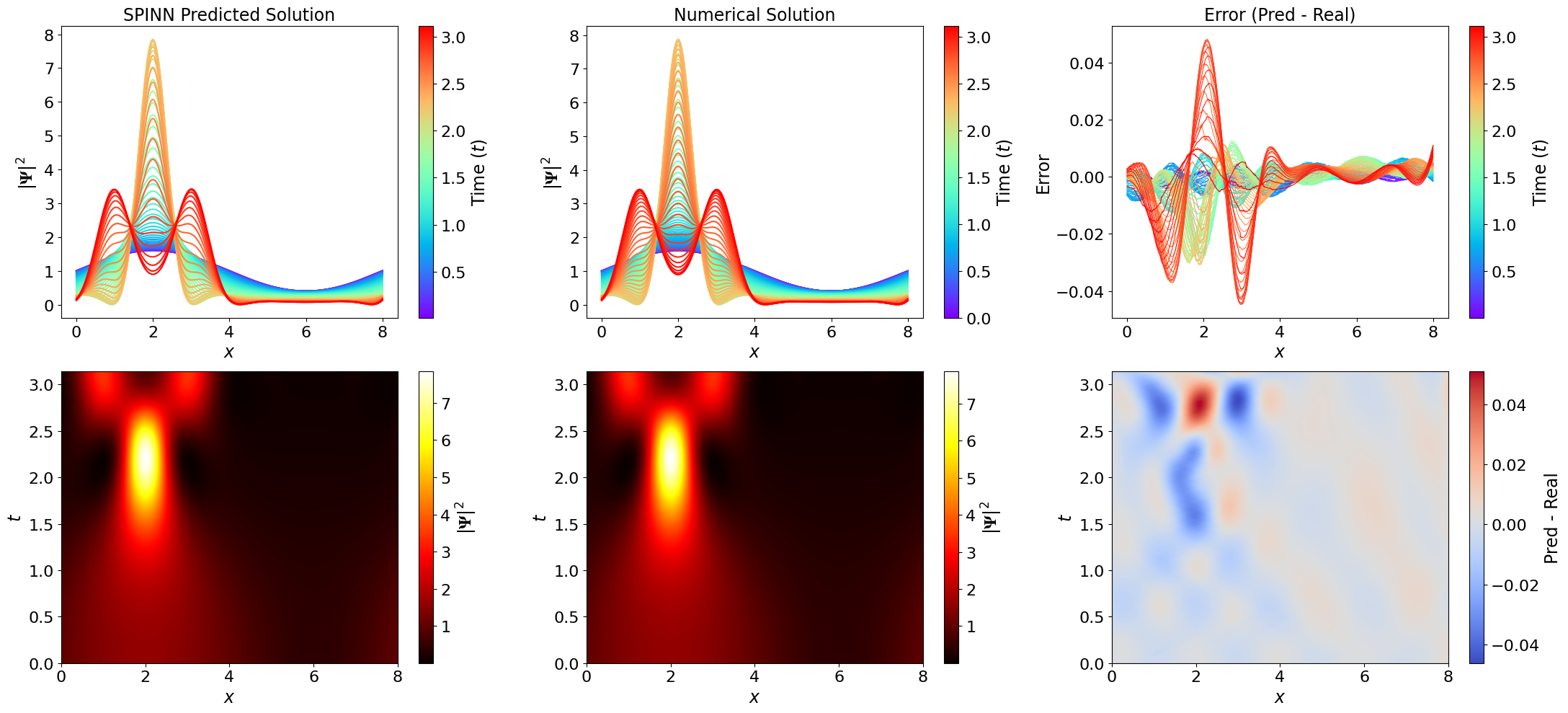}

\caption{Time evolution of the 1-D density distribution under the effect of the gravitational potential, predicted with SPINN: In the top row, density distributions are shown at fixed time frames indicated by the color gradient, with spatial coordinate on the $x$-axis and density on the $y$-axis. From left to right, the plots show the SPINN prediction, the corresponding numerical solution with Spectral Method (described in Appendix \ref{appendix_A}), and the error between them. In the bottom row, the same evolution is shown in a 2-D perspective using heatmaps: the $x$-axis represents the spatial coordinate, the $y$-axis denotes time, and the density magnitude (or error) is represented through a color gradient.}

    \label{Fig2}
\end{figure*}
Before we introduce the set-up for the results, it is necessary to discuss the inherent scale in the SP equations and the corresponding units relevant for this work. The SP equations (\ref{SE_re}-\ref{PE_re}) remain invariant under the following scaling transformation:
\begin{equation*}
    \{t,x\} \mapsto  \{\alpha t,\beta x\}
\end{equation*}
if and only if 
\begin{equation*}
    \{V,\lambda\} \mapsto  \{\beta^{-2} V, \beta^{-2} \alpha \lambda\}
\end{equation*}
The parameter $\lambda$ naturally emerges as a fundamental scale within the system, as it governs the scaling of both spatial and temporal coordinates. As a result, two systems that differ in domain size or timescale (i.e., different choices of $\beta$, and $\alpha$) but satisfy the same rescaled SP equations, can exhibit qualitatively different dynamics unless 
$\lambda$ is appropriately scaled. This highlights the need to fix $\lambda$ or work in completely dimensionless units, when comparing solutions or setting up numerical experiments, to account for how changes in box size or evolution time implicitly alter the dynamics through $\lambda$.
\section{Results and Discussion}\label{Res}
\subsection{1D and 3D Inital and Boundary Conditions}
We choose a one dimensional length of L = 8. Following the Section \ref{GE} on Governing Equations, we use arbitrary units for both the spatial and temporal dimensions. We pick the following well-known standard test case as the initial condition in our 1D setting, for our results to be comparable with \citet{2021_mocz}, and \citet{2024_Cappelli}:
\begin{equation}
    \Psi(x,0) = \sqrt{1.0 + 0.6\sin\left(\frac{\pi x}{4}\right)}
\end{equation}
Which represents one of the Fourier components of the Gaussian initial condition often used in cosmological CDM simulations. This can be easily extended to 3D set-up as follows:
\begin{equation}
    \Psi(\boldsymbol{x},0) = \sqrt{1.0 + 0.6\sin\left(\frac{\pi x}{4}\right)\sin\left(\frac{\pi y}{4}\right)\sin\left(\frac{\pi z}{4}\right)}
\end{equation}
Periodic boundary conditions (BCs) are enforced for $\Psi$ and V. For both the gravitational potential V and the wavefunction $\Psi$, periodicity must also be imposed on their first spatial derivatives given that they are governed by equations involving second-order partial derivatives. For example, along the $x$-direction
\begin{align*}
    \Psi(x = 0,y,z,t) = \Psi(x=L,y,z,t) \\
    \partial_x\Psi(x= 0,y,z,t)  =  \partial_x\Psi(x=L,y,z,t)
\end{align*}
\begin{align*}
    V(x = 0,y,z,t) = V(x=L,y,z,t) \\
     \partial_xV(x= 0,y,z,t)  =  \partial_xV(x=L,y,z,t)
\end{align*}
with $L$ being the domain length in the $x$-direction. In the same way, it is implemented along all the independent dimensions of the domain. This periodicity condition means that particles exiting any boundary of the domain re-enter from the opposite side.

\begin{figure}
    \includegraphics[width=0.85\linewidth]{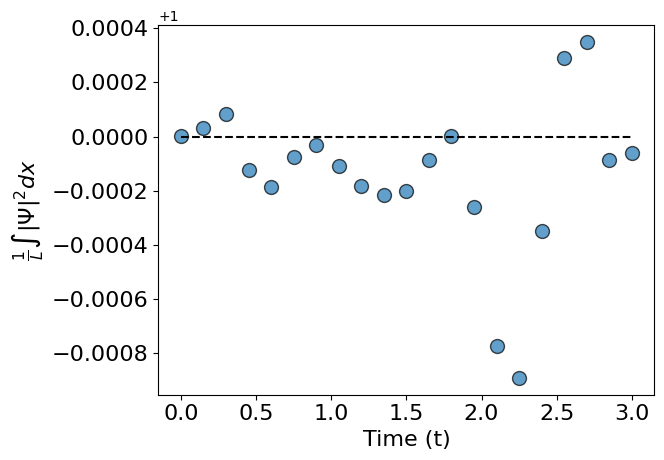}
    \caption{Evaluation of a key metric, mass conservation (or equivalently normalization of the state): The plot shows the normalization at fixed time frames, with time on $x$-axis and $|\boldsymbol{\Psi}|^2$ integrated over the spatial coordinate on the $y$-axis.}
    \label{Fig3}
\end{figure}

\subsection{Interpretation and Checks}
\begin{figure*}
\centering
\begin{tabular}{ccc}
\includegraphics[width=0.3\linewidth]{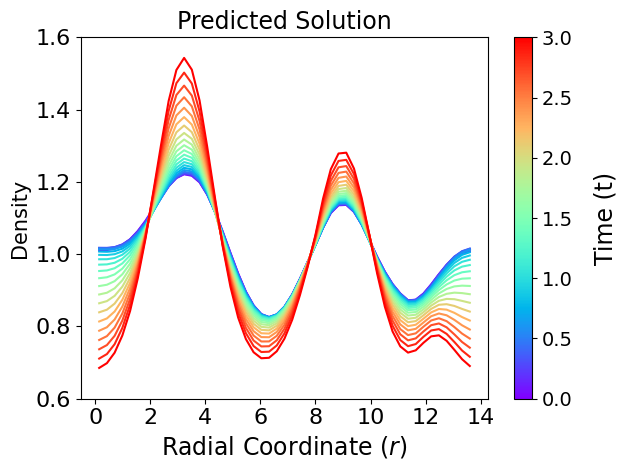} &
\includegraphics[width=0.3\linewidth]{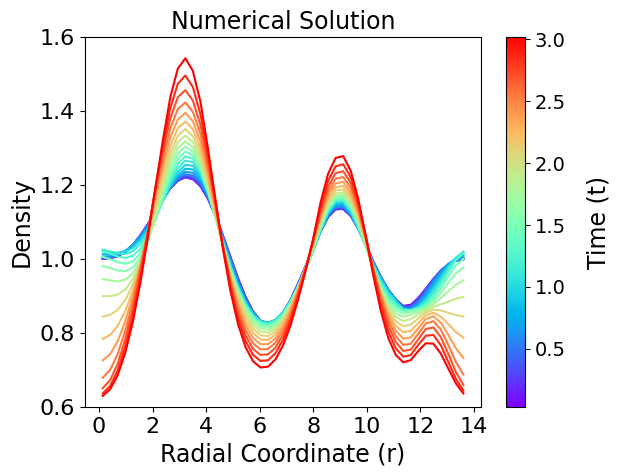} &
\includegraphics[width=0.3\linewidth]{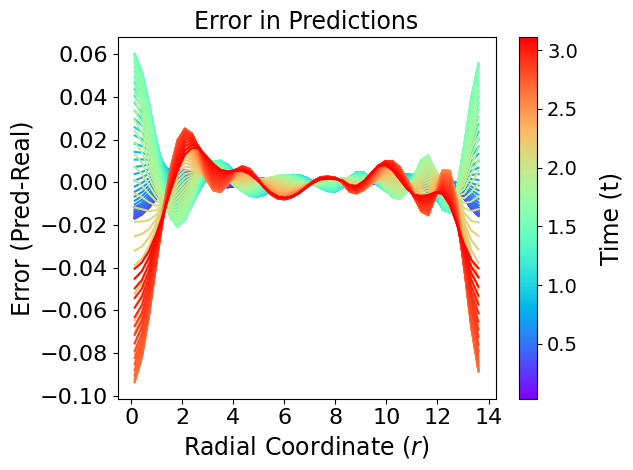}
\end{tabular}
\caption{Average radial density at different time frames from the  (i) SPINN Predictions (left), (ii) Numerical Solution with spectral second-order unitary numerical scheme - See appendix \ref{appendix_A} (middle), and (iii) Error between the two (right). Plots show evolution of average radial density (or error in its prediction) as a function of the radial coordinate (defined as $r = \sqrt{x^2+y^2+z^2}$), in particular the growth of overdensities over time.}
\label{Fig4}
\end{figure*}
Fig.(\ref{Fig2}) presents the time evolution of an initial sinusoidal perturbation governed by the Schrödinger–Poisson equations (\ref{SE_re}–\ref{PE_re}) over the interval $t = 0$ to $t=3$, for the chosen coupling parameter 
$\lambda = 1$. The top row of the figure displays the predicted density evolution by SPINN (left), the corresponding numerical solution obtained via the spectral method (center; see Appendix~\ref{appendix_A}), and the absolute error between the two (right). The density profiles are shown at selected time frames, with color gradients indicating temporal progression.
The lower panel offers a 1+1D visualization of the same dynamics as a surface plot: the horizontal axis corresponds to space, the vertical axis to time, and the color encodes the density magnitude (or error, in the third subpanel). Physically, the collapse and subsequent splitting of the wavefunction can be attributed to the self-interacting gravitational potential, whose strength is modulated by the parameter 
$\lambda$. The error between SPINN and the spectral solution remains generally low throughout the evolution, with deviations primarily occurring near the sharp region of densities at later times. These discrepancies are likely due to weak enforcement of boundary conditions and the presence of competing terms in the multi-component loss function, leading to accumulation of error over time.
 
Since we impose periodic boundary conditions, the total mass—represented by the spatially averaged density ($\bar\rho$)—should be conserved. In our case, this implies that the wavefunction should retain the same normalization throughout its evolution. The plot of Fig.(\ref{Fig3}) shows the wavefunction normalization over time, which remains approximately one, confirming that SPINN successful learns conservation of mass and probability. 

Furthermore, Fig.(\ref{Fig4}) shows the time evolution of the average density plotted against the radial coordinate (defined as $r = \sqrt{x^2+y^2+z^2}$), for the 3D extension of the initial sinusoidal perturbation as stated in the previous subsection. It can be seen that the overdensities collapse as the time evolves. The numerical solutions is also depicted in the same figure, showing a very good match except at the corner regions (i.e, the boundaries of the radial coordinate). This may be related to the smoothness of the sine activation function used in SPINN that the predicted solution varies gradually at the corners, unlike the numerical solution due to the inherent error associated to the method.
\begin{figure*}
    \centering
    \includegraphics[width =0.3\linewidth]{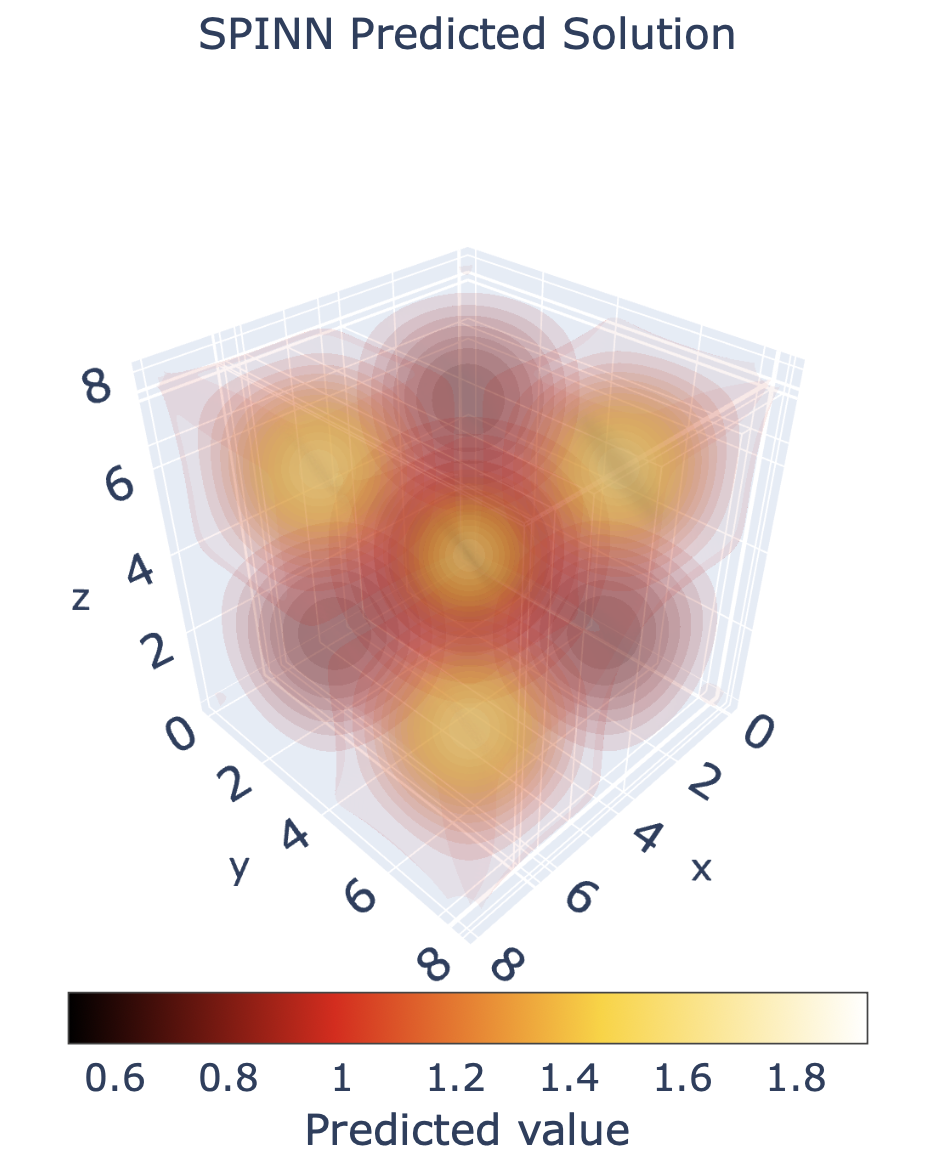}
    \includegraphics[width =0.36\linewidth]{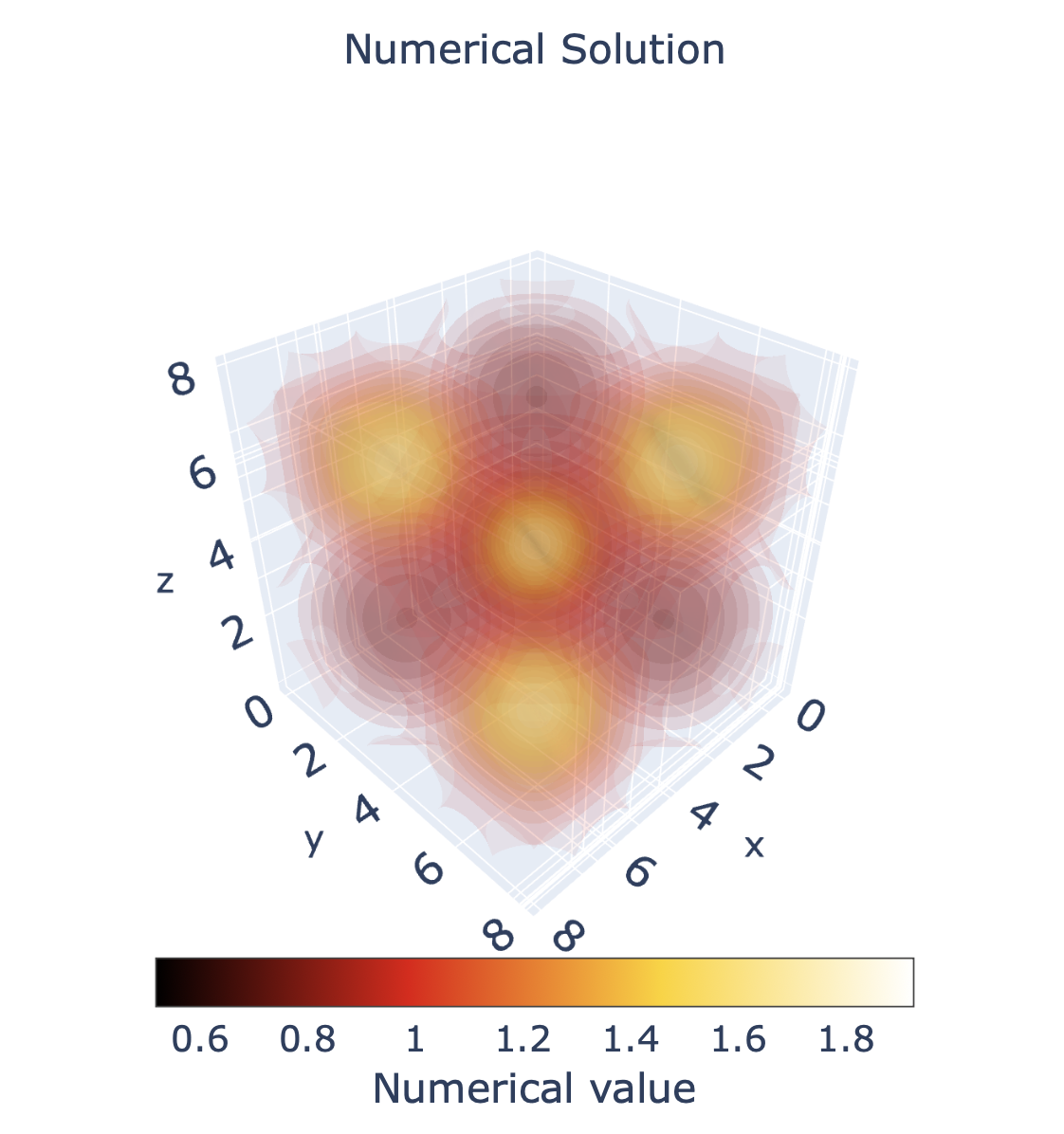}
    \includegraphics[width =0.325\linewidth]{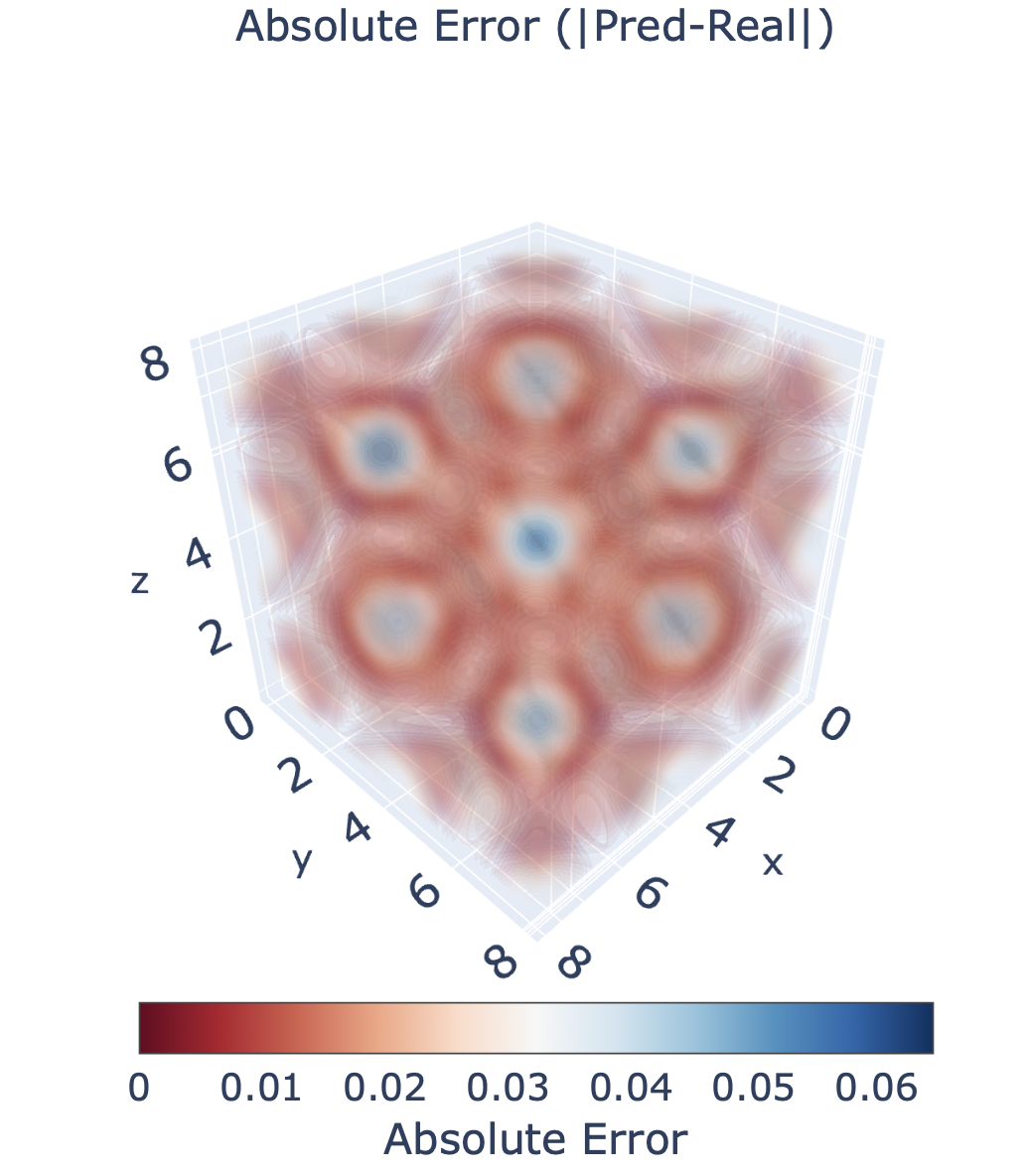}
    \caption{Evaluation procedure for randomly chosen density cube at a fixed time frame. From left to right, the first panel shows a density cube at randomly chosen $t$ ($t = 1.0367$) from the numerically computed solution of SP equations in 3D. The second panel exhibits the same as predicted by our trained SPINN model. The third panel  shows the absolute error between the predicted and the actual volumes.}
    \label{fig:55}
\end{figure*}

\begin{figure*}
\centering
\hspace*{-1.0cm}
\begin{tabular}{cccc}
\includegraphics[width=5.4cm,height = 4.8cm]{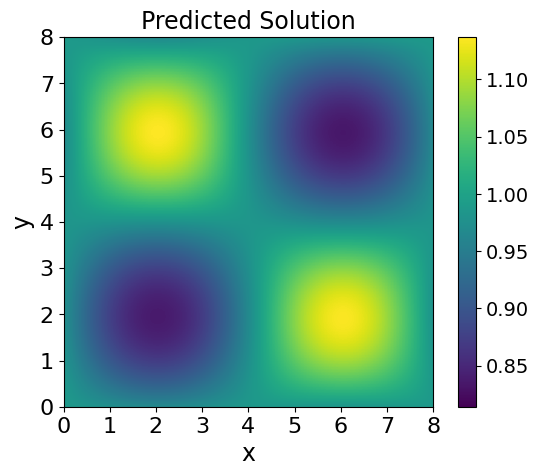} & \includegraphics[width=5.4cm,height = 4.8cm]{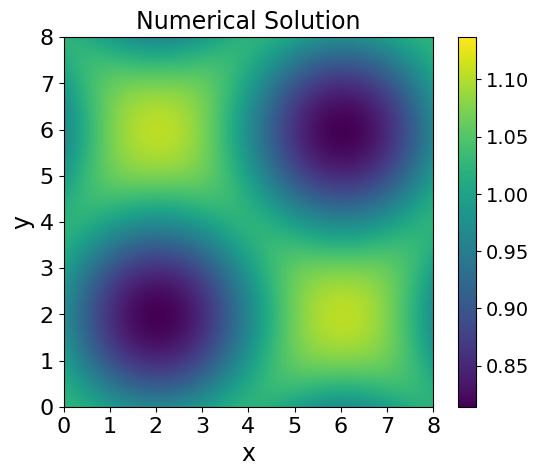}&
\includegraphics[height = 4.8cm,width=5.4cm]{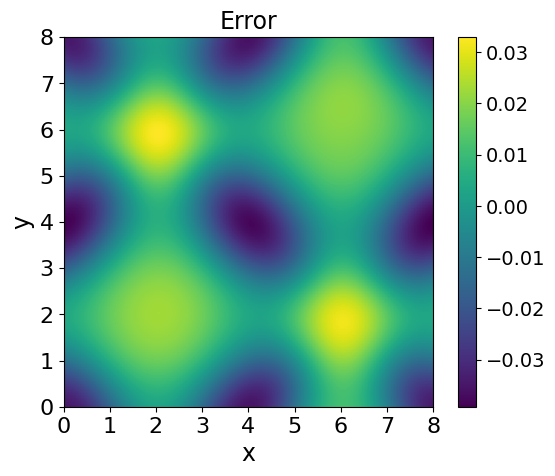}
\\
\end{tabular}
\caption{Evaluation procedure for randomly chosen $x-y$ density slice at a fixed $z$ ($z = 0.3125$) and time frame ($t = 1.0367$). From left to right, the first panel shows an $x-y$ slice at randomly chosen $z$ and $t$ as predicted by our trained SPINN model in 3D. The second panel exhibits the same slice obtained from the numerically computed solution of SP equations in 3D.} The third panel  shows the error between the predicted and the actual slice.
\label{Fig5} 
\end{figure*}
To further validate the predicted solution,
we compare both a full density cube and a 2D density slice prediction randomly chosen at a fixed $z$ and a time frame. 
Fig.(\ref{fig:55}) shows a density cube at a fixed time frame (
$t=1.0367$), comparing the SPINN prediction (left) with the numerically computed solution (middle), followed by the absolute error between the two (right). Similarly, Fig.(\ref{Fig5}) shows a corresponding 
$x-y$ density slice obtained from the SPINN and the numerical solution, along with the absolute error associated between the two, taken at a randomly chosen $z$-location (
$z=0.3125$) at the same time frame. In both cases, the absolute error remains close to zero across the domain, confirming that SPINN accurately reconstructs the 3D density field and its lower-dimensional representations.
\subsection{Extrapolation in time}
\citet{2020_Kim} demonstrate that PINNs can be trained such that they can be efficiently used for extrapolation, and thus can be used for predictions even for the time range for which they are not trained for. This enables PINNs to be utilized for various other applications including but not limited to surrogate modelling, uncertainty quantification \citep[e.g. ][]{2019_Zhu,2020_Sun,2021_Hoffer} as well as inverse problems \citep[See ][]{2021_Yang}. We illustrate this ability for SPINNs as well. In particular, Fig. 7 presents 2D heatmaps of the numerical and predicted solutions over a temporal range from $t = 0$ to slightly beyond $t = 4$, extending past the edge of the training domain ($t\approx 3$).
It can be observed that the solutions are in good agreement uptil $t = 4$ at which point it starts to deviate from the numerical solution. This extrapolation has been obtained using standard training methods unlike the \citet{2020_Kim} where novel methods have been adopted to extend these extrapolation intervals. This shows the potential of SPINN to extrapolate the solutions obtained and room for improvements in its future versions with better training approaches.
\begin{figure}
    \centering
    \includegraphics[width=\linewidth]{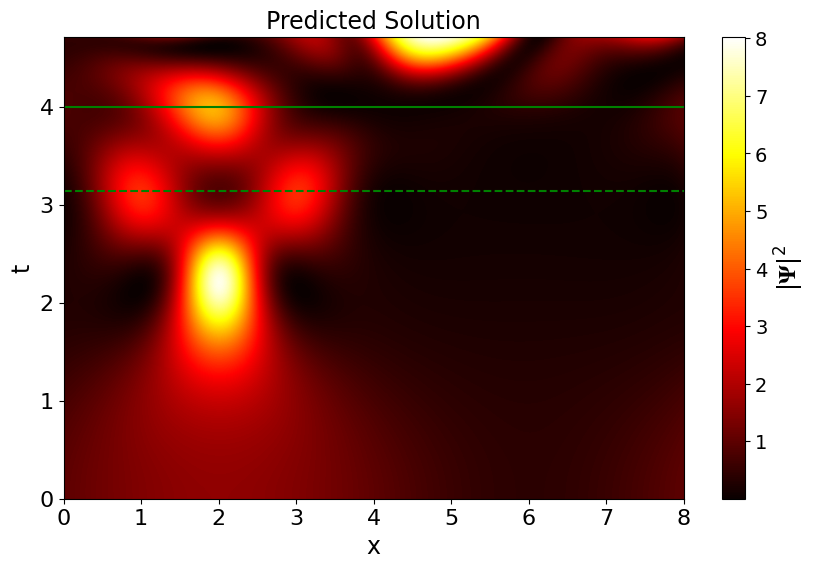}
    \includegraphics[width=\linewidth]{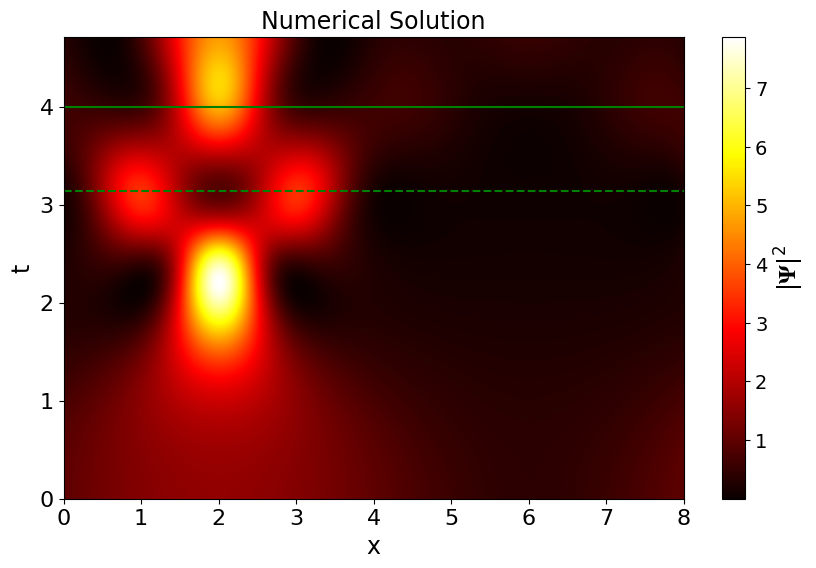}
    \caption{Extrapolating the SPINN predictions to outside the training time units: the top panel shows the predicted solution to approximately t = 4 and the bottom panel depicts the numerical solution for the same time units, both are represented by the 2D heatplot with spatial coordinate on x and time coordinate on y. The color gradient represents the density. The dashed line in both the plots represents the training time of the model, while the solid line marks roughly the point beyond which the predicted and numerical solutions begin to diverge.}
    \label{Fig6}
\end{figure}
\subsection{Wavefunction normalization as an additional constraint}
Some works \citep[e.g.][]{2024_Brevi} advocate the use of an additional normalization constraint as necessary for optimizing the PINNs used to solve the Schr\"odinger equation. In our work the SPINN is able to learn the optimal solution without such a  normalization constraint. In fact, the normalization of the wavefunction automatically follows as the network correctly learns the physics of the system, i.e. by minimizing the PDE residuals. This is expected, as any wavefunction obeying Schr\"odinger's equation should also satisfy the continuity equation. 

\begin{figure*}
\centering
\begin{tabular}{ccc}
\includegraphics[width=5.9cm]{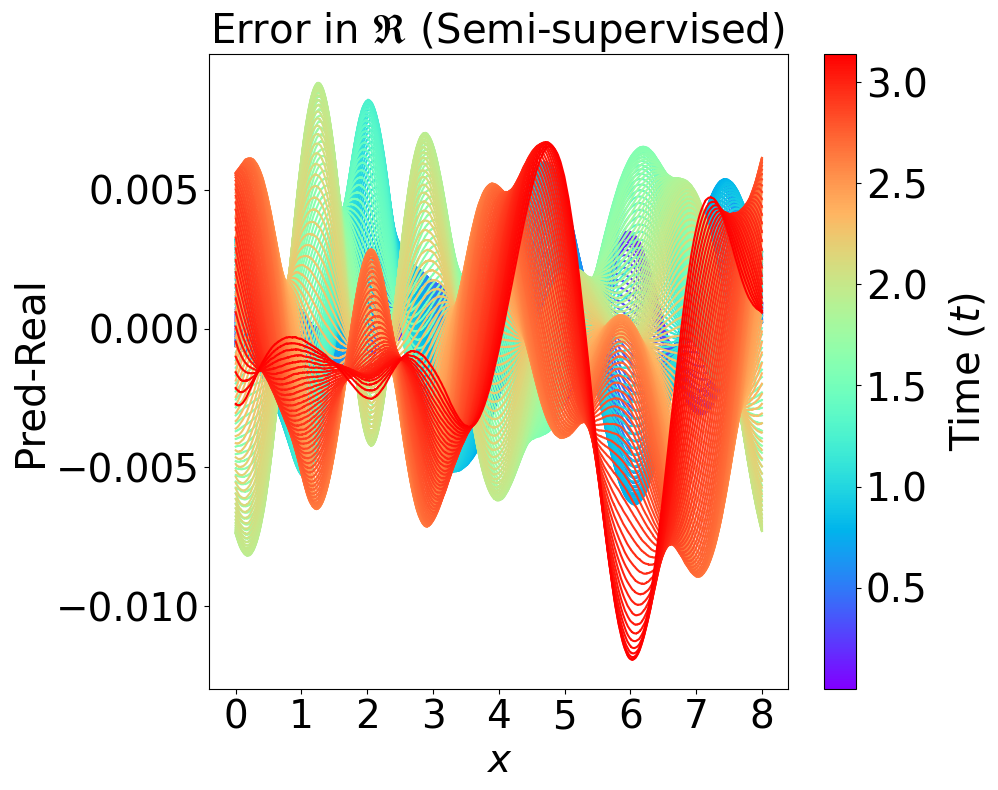}&
\includegraphics[width=5.9cm]{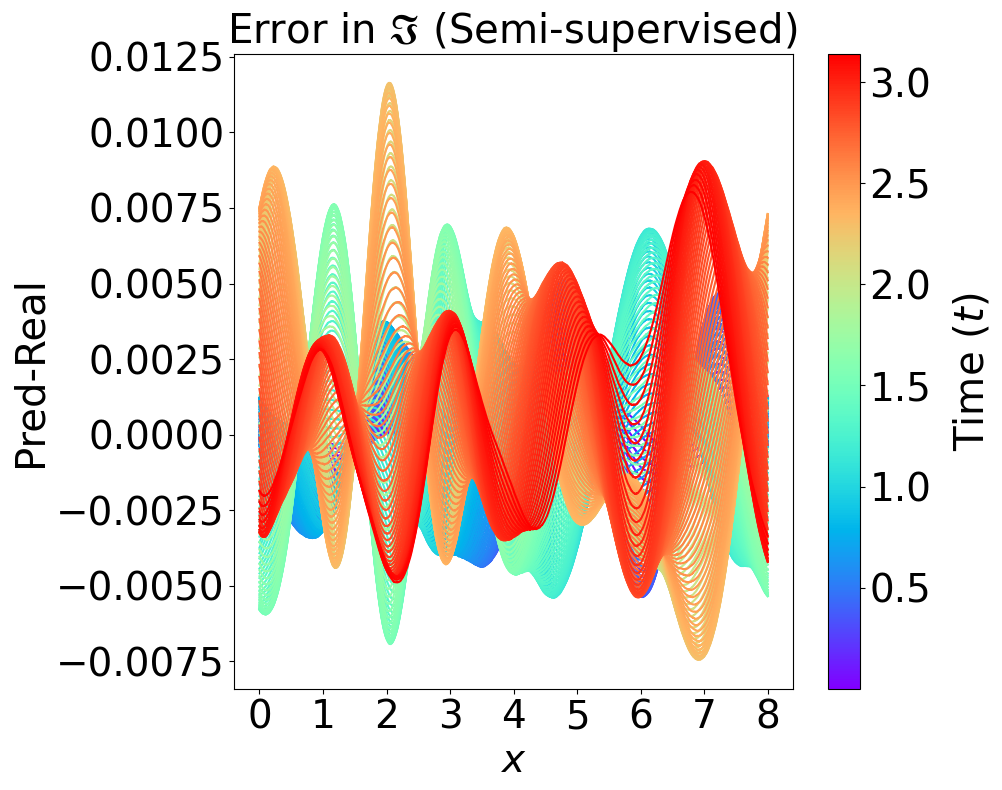}
&\includegraphics[width=5.9cm]{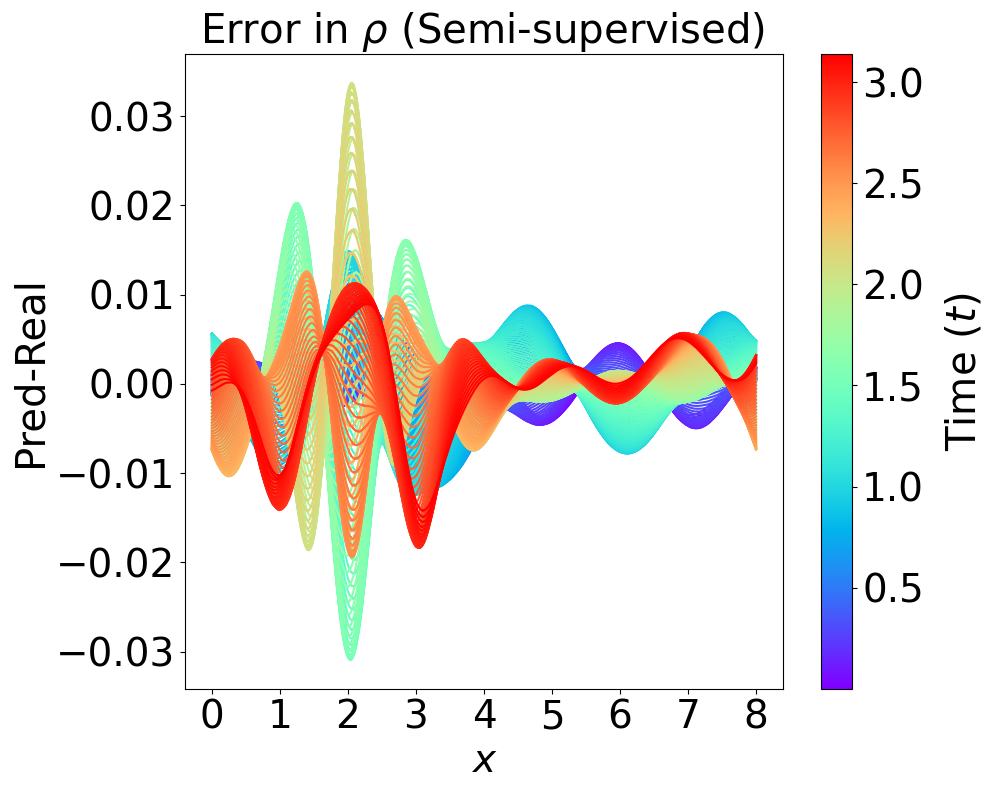}
\end{tabular}
\caption{Semi-Supervised Learning: Errors in $\rho,\mathfrak{R},\mathfrak{I}$ are shown here with respect to the analytical solution (See appendix \ref{appendix_A})}
\label{Fig_last}
\end{figure*}
\begin{figure*}
\centering
\begin{tabular}{ccc}
\includegraphics[width=5.9cm]{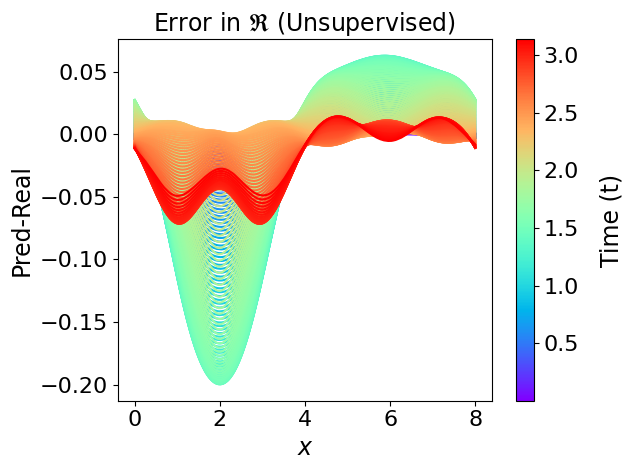}&
\includegraphics[width=5.9cm]{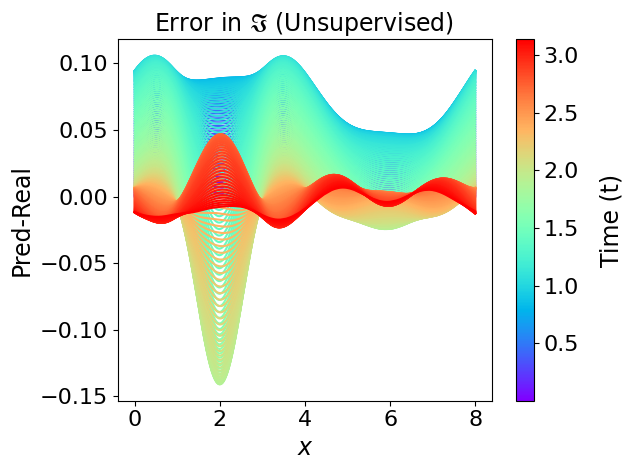}
&\includegraphics[width=5.9cm]{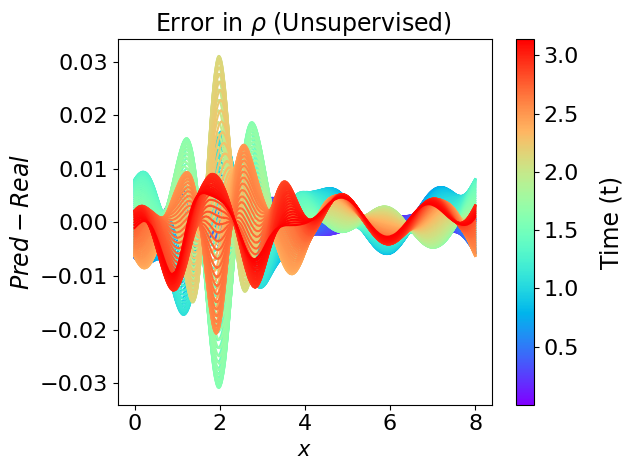}
\end{tabular}
\caption{Unsupervised Learning: Errors in $\rho,\mathfrak{R},\mathfrak{I}$ are shown here with respect to the analytical solution (See appendix \ref{appendix_A})}
\label{Fig_unsup}
\end{figure*}
\subsection{A Note on Predicting Potential as an Additional Output}\label{Aux}
In our approach, we predict both the real and imaginary components of the wavefunction, along with the potential, using a neural network. A natural question arises: why predict the potential directly from the network when it can be computed using a Poisson solver?
To investigate this, we implemented a Physics-Informed Neural Network (PINN) that predicts only the real and imaginary parts of the wavefunction while solving for the potential using the Fourier transform. The results of this approach are presented in Appendix \ref{av_app}. However, this method necessitates a trade-off in spatial resolution. Training such a PINN at high resolutions is computationally expensive, both in terms of time and memory.  Specifically, computing the Fourier transform for an $N \times N$ grid requires $\mathcal{O}(N^2logN)$ operations and approximately $\mathcal{O}(N^2)$ memory, which can quickly become prohibitive for large grids. As a result, memory overflow errors often occur, forcing us to lower the resolution to accommodate computational constraints.

By contrast, predicting the potential as an additional output eliminates these computational bottlenecks, enabling faster convergence and maintaining high accuracy without the need to compromise resolution.
\subsection{Semi-Supervised Learning}\label{SS_sec}
We have thus far presented results using unsupervised learning for PINNs. However, the scalability of this approach to longer time spans and larger domains remains uncertain. For FDM simulations, a potential strategy is to leverage large-scale CDM simulations via semi-supervised learning and refine small-scale features using physics constraints. In this work, we briefly investigate this approach in 1D by training semi-supervised networks and comparing their accuracy to unsupervised PINN training. To ensure a fair comparison, we employ neural networks with identical architectures in both cases—each consisting of 7 hidden layers with 32 neurons per layer—and train them using the same optimization strategies and loss formulations as detailed in Section \ref{method}. The only distinction lies in the semi-supervised case, where a limited set of simulation data is incorporated into the loss function to provide an additional data-driven component. 

Fig.(\ref{Fig_last}) shows the results of the semi-supervised model, while Fig.(\ref{Fig_unsup}) displays predictions from the corresponding hard-conditioned unsupervised PINN, trained using the same $7\times32$ architecture. Clearly, semi-supervised training yields improved accuracy in the final results, particularly in capturing the real and imaginary components of the solution, which together encode the wavefunction’s phase—a critical quantity for representing wave-like features in FDM. . These results underscore the potential of incorporating limited data into physics-informed networks to enhance generalization. Further details on the implementation and training behavior of the hard-conditioned model can be found in Appendix~\ref{hc_app}.
\section{Conclusion}\label{conc}
Emerging Physics-Informed Neural Networks (PINNs) 
have the potential to revolutionize cosmological simulations, particularly in the context of Fuzzy Dark Matter (FDM). Previous works have been constrained by limited resolution, restricting simulations to small system sizes and high redshifts. PINN-based ML techniques may provide a path towards overcoming these limitations, with GPU hardware potentially enabling faster implementations and mitigating the exponential increase in computational cost. 

This work demosntrates a first step in this direction with the SPINN, a PINN-based framework for efficiently solving the non-linear Schrödinger-Poisson equations in FDM cosmological simulations. We demonstrate its accuracy by evolving an initial sinusoidal perturbation—a standard test case for Vlasov-Poisson equations—in both 1D and 3D. Results show accurate density predictions, mass conservation, and a close match with the numerical solution from a second-order spectral scheme \citep{Mocz_2017_BECDM}. An additional advantage of SPINN is its ability to extrapolate beyond the training time, though further development is needed for longer-scale extrapolations.

A key challenge for scaling this framework to full cosmological simulations is the spatiotemporal domain size and complexity of initial conditions. While this work represents an important first step, the training time was not sufficiently long, nor the initial condition complex enough, to fully assess the scalability of the framework for longer times or more intricate scenarios, where spectral bias may affect performance \citep{2018_Rahaman}. However, recent advancements suggest solutions, such as multi-level domain decomposition-based PINN architectures \citep{2024_Dolean} and Fourier embedding techniques for addressing spectral bias \citep{2020_Tancik}.

It may be possible to improve the accuracy of unsupervised PINNs using advanced architectures such as Pinnsformer \citep{2023_Pinnsf}, PIKANs \citep{2024_Shuai}, and Pinnmamba \citep{2025_Xu}.
However these networks will still be difficult to optimize for the large spatiotemporal domains necessary for cosmological simulations. But our results clearly show that blending the simulations with physics constraints, is a viable way to proceed for cosmological simulations especially given that we already have publicly available large scale CDM simulations, which could be used as data constraint and then FDM evolution equations can be used to further supplement them to ``paint-in" the small scale features.

While the current framework shows strong potential for modeling the Schrödinger-Poisson system over limited time frames, we believe it can be scaled to longer durations and more complex conditions with the right combination of techniques. In conclusion, PINNs, when combined with generative models, may offer a promising avenue for developing large-scale FDM cosmological simulations, which can eventually be compared to existing CDM simulations. Future work will focus on extending SPINN to large spatial and temporal domains, using cosmological initial conditions at high redshifts.
\section*{Acknowledgments}
AK and ET acknowledge financial support from the SNSF under the Starting Grant project Deep Waves (218396). The authors thank Nicolas Cerardi for his valuable discussions during the project. This work was supported by EPFL through the use of the facilities of its Scientific IT and Application Support Center (SCITAS). The use of facilities of the Swiss National Supercomputing Centre (CSCS) is gratefully acknowledged.
\section*{Code Availability}
All models are implemented in PyTorch \citep{2019_Paszke}, and are trained
separately on single GPU. Code for this work (along with the scripts to produce the plots) is publicly available on Github at the following address: (\href{https://github.com/mishraashu6566/DeepWaves-SPINN}{DeepWaves-SPINN}). .
\bibliography{mybib}{}

\appendix 
\section{ Numerical Algorithm for Solving the Schrödinger-Poisson Equations}\label{appendix_A}

In this appendix, we detail the numerical method employed to solve the Schrödinger-Poisson (SP) equations, following the 2nd Order Unitary Spectral Method as described in \citet{Mocz_2017_BECDM}. This method utilizes spectral techniques and Fast Fourier Transforms (FFT) for computational efficiency.

The SP equations (\ref{SE_re}-\ref{PE_re}) govern the evolution of a quantum fluid system, where the wavefunction \( \Psi \) is influenced by gravitational interactions. The numerical algorithm employs a time-splitting spectral approach, involving half-step potential updates (‘Kick’), full-step kinetic evolution (‘Drift’), and iterative potential updates.

\begin{algorithm}[H]
\caption{2nd Order Unitary Spectral Method for Schrödinger-Poisson Equations}
\begin{algorithmic}[1]
\State Initialize wavefunction \( \Psi \) and compute initial potential \( V \)
\While {time \( t < t_{\text{final}} \)}
    \State Compute gravitational potential:
    \[
    V = \text{IFFT} \left( \frac{-1}{k^2} \text{FFT} \left( 4\pi Gm (|\Psi|^2 - |\Psi_0|^2) \right) \right)
    \]
    \State Apply half-step potential ‘Kick’:
    \[
    \Psi \gets \exp \left[ -i \left( \frac{m}{\hbar} \right) \left( \frac{\Delta t}{2} V \right) \right] \Psi
    \]
    \State Apply full-step kinetic ‘Drift’ in Fourier space:
    \[
    \Psi \gets \text{IFFT} \left( \exp \left[ -i \Delta t \left( \frac{\hbar}{m} k^2 \right)/2 \right] \text{FFT}(\Psi) \right)
    \]
    \State Recompute gravitational potential:
    \[
    V = \text{IFFT} \left( \frac{-1}{k^2} \text{FFT} \left( 4\pi Gm (|\Psi|^2 - |\Psi_0|^2) \right) \right)
    \]
    \State Apply another half-step potential ‘Kick’:
    \[
    \Psi \gets \exp \left[ -i \left( \frac{m}{\hbar} \right) \left( \frac{\Delta t}{2} V \right) \right] \Psi
    \]
    \State Advance time step: \( t \gets t + \Delta t \)
\EndWhile
\end{algorithmic}
\end{algorithm}

The described numerical scheme preserves second-order accuracy and ensures unitary time evolution. The spectral approach efficiently handles spatial derivatives and gravitational interactions, making it well-suited for simulations of Bose-Einstein condensates and wave-like dark matter models. This method is particularly effective in capturing the quantum mechanical effects in astrophysical and condensed matter applications.

\section{ SPINN with Madelung Formalism}\label{appendix_B}
The Madelung formalism for FDM simulations is often preferred as it is straightforward to incorporate it into the existing hydrodynamical codes \citep{2018_Nori,2024_Kunkel}. However, the most challenging aspect of simulating the Madelung equations is to model the quantum pressure term which is ill-defined in regions of vanishing densities. In the following, we compare PINNs' results obtained using the Madelung and Schrödinger-Poisson (SP) formalisms. We start with the Madelung formalism expressed in terms of density and phase with $\lambda = 1$:

\begin{align}
 \label{B1}   \partial_t \rho + \nabla \cdot (\rho \nabla S) = 0 \\
\label{B2}    \partial_t S + \frac{1}{2} (\nabla S)^2 + V -\frac{1}{2}\frac{\nabla^2 \sqrt\rho}{\sqrt\rho} = 0\\
\label{B3}    \nabla^2 V - (\rho-1) = 0
\end{align}
We note that the Madelung formalism assumes the density $\rho$ is strictly positive ($\rho>0$), an assumption that is not generally valid. In many physical situations, including the ones we study, the density can vanish locally, and the validity of the Madelung representation in these regions remains unclear. To address practical challenges while training the network, we slightly modify the Madelung equations by rewriting the quantum pressure term to eliminate explicit divisions by 
$\rho$, which otherwise introduce large numerical instabilities and exploding gradients during network optimization.
Writing Eq.(\ref{B2}) in 1D, and expanding the quantum pressure term using the chain rule of derivatives, we get the following expression:
   $$\rho^{-1/2}\frac{\partial^2}{\partial x^2}\Big(\sqrt{\rho} \Big) = -\frac{1}{4} \rho^{-2} \Big(\frac{d\rho}{dx}\Big)^2 + \frac{1}{2} \rho^{-1} \frac{d^2\rho}{dx^2}$$ 
Substituting this back in Eq. (\ref{B2}) and multiplying both sides by $\rho^2$, we get
 $$\boxed{ 2\rho^2\Big[ \partial_t S + \frac{1}{2}( \partial_x S)^2 + V  \Big] - \Big[-\frac{1}{4} \Big(\frac{d\rho}{dx}\Big)^2 + \frac{1}{2} \rho \frac{d^2\rho}{dx^2} \Big] = 0}$$
\begin{figure*}
    \centering
    \includegraphics[width=\linewidth]{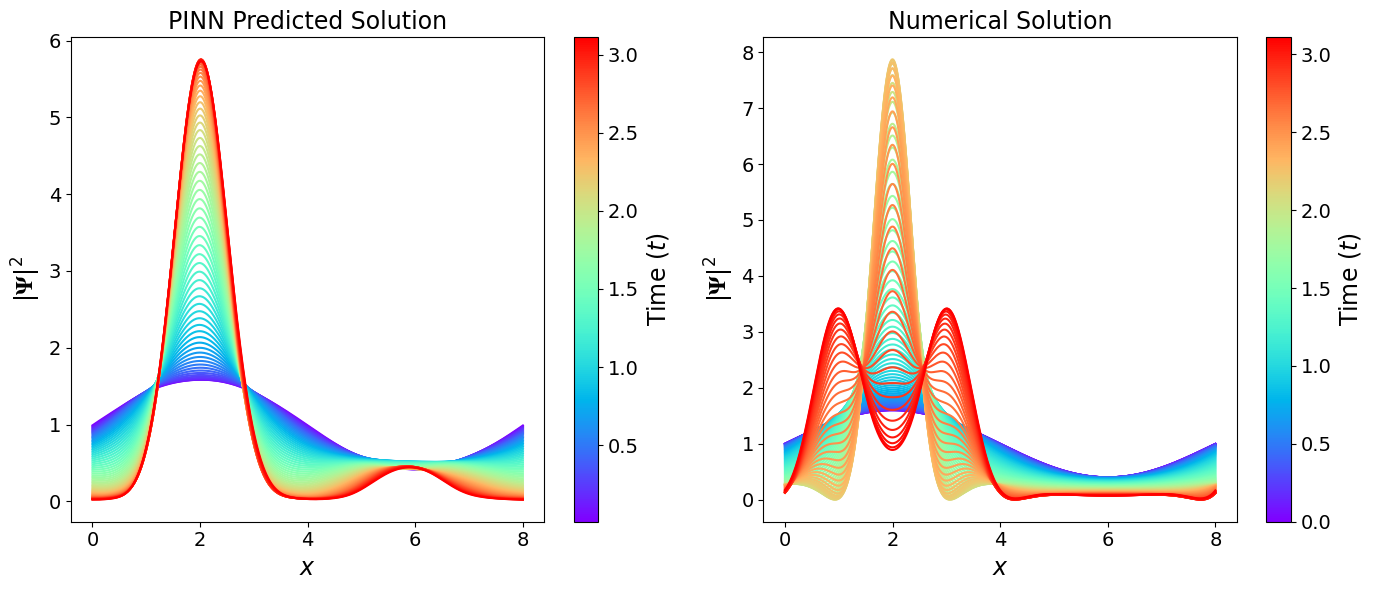}
    \caption{Time Evolution of Density from PINN with Madelung formalism: 
    density distributions are shown at fixed time frames indicated by the color gradient, with spatial coordinate on the $x$-axis and density on the $y$-axis. From left to right, the plots show the PINN prediction and the corresponding numerical solution with Spectral Method (described in Appendix \ref{appendix_A}).}
    \label{Fig17}
\end{figure*}

This form of Eq. (\ref{B2}) is used hereafter. We replace the standard Schrödinger-Poisson Equations (\ref{SE_re}-\ref{PE_re}) with the Madelung formalism, specifically using Equations (\ref{B1},\ref{B2},\& \ref{B3})  as the physics constraint during training. Under this formulation, the network is tasked to predict the density $\rho$, the phase $S$, and the potential V.  We keep the neural network architecture, optimizer, and all hyperparameters identical to the setup used for direct SP Equations training to ensure a fair comparison. This way, any differences in performance can be directly attributed to the change in physical formulation, rather than changes in network capacity or training strategy.

Fig.(\ref{Fig17}) shows the 1D density evolution obtained from SPINN using fluid or Madelung Formalism, with spatial coordinate on $x$-axis and the density on $y$-axis, with time represented by the color gradient. It depicts that the network ends up learning the collapse behaviour but cannot capture the wave-like nature of the coupled potential as we no longer observe the splitting of the density peak into two. This may be attributed to the fact that the Madelung Formalism cannot capture the wave features as the SP Equations and Madelung Equations are not strictly equivalent \citep{2018_Zhang}. \citet{Wallstrom_1994} proved that a quantization condition $\oint \nabla S  \cdot dl = 2\pi j$ ($j\in\mathbb{Z}$ and L is any closed loop) is required to recover the SP Equations. 

\section{Additional Tests Results}\label{appendix_c}
\subsection{Hard-Conditioned Unsupervised PINNs}\label{hc_app}
As stated in Section~\ref{SS_sec}, we provide the implementation details for the hard-conditioned PINN in the 1D setting. We define a neural network $NN(X;\theta)$ which approximates $\mathfrak{R}$, $\mathfrak{I}$, and $V$, where any of these outputs is represented by u. The approximation is then given by:
\begin{align}
    \hat{u}(X;\theta) = u_0 + t \, NN(X;\theta)
\end{align}
where $(\cdot)_{\theta}$ indicates a PINN approximation realized with a network represented by the set of parameters $\theta$. Here, $u_0$ represents the initial condition for the corresponding quantity. Thus, the PINN-framework can be interpreted as modeling the displacement evolution from the initial condition. By construction, this method inherently enforces the initial condition, overcoming a potential drawback of using an initial loss term in PINN training. In the conventional approach, the initial condition is only weakly imposed, which may lead to inconsistencies in the learned solution. Furthermore, recent studies have both theoretically and empirically demonstrated that PINN optimization can suffer from stiffness, making it difficult to solve and, in some cases, preventing convergence altogether due to competing loss terms \citep{2020_Wang, 2021_Wang, 2020_Sun}. The SPINN results with this hard-conditioning, and identical architecture as the semi-supervised case described in Section \ref{SS_sec} are shown in Fig. (\ref{Fig_unsup}). Clearly, the semi-supervised results shown in Fig. (\ref{Fig_last}), have lower  
errors in terms of predictions of real and imaginary components of the wave function than that of the hard-conditioned unsupervised case. However, the error in densities are comparable. We also note that the hard-conditioned unsupervised PINNs perform better than the usual unsupervised PINNs which can be seen from Fig.(\ref{Fig_unsup}) and Fig. (\ref{Fig2}).

\subsection{Training PINN without predicting potential as an additional output}\label{av_app}

Here we present the detailed results of the experiment discussed in Subsection \ref{Aux}. In this setup, we train a PINN to predict only the real and imaginary parts of the wavefunction \(\psi\) using the same network set-up (i.e. same architecture, optimizer, and all hyperparameters as used in SPINN for 1D case), while computing the potential \(V\) externally by solving the Poisson equation using the Fourier transform at each training step.

Specifically, after the network predicts \(\text{Re}(\psi)\) and \(\text{Im}(\psi)\), we evaluate the density \(|\psi|^2\) and solve for \(V\) by applying a Fourier-based solver to the equation \(\nabla^2 V = |\psi|^2-1\). This approach avoids making \(V\) an explicit network output and treats it instead as a dependent variable. The network loss is constructed using the predicted wavefunction components and the externally computed potential.

\begin{figure}
    \centering
    \includegraphics[width=\linewidth]{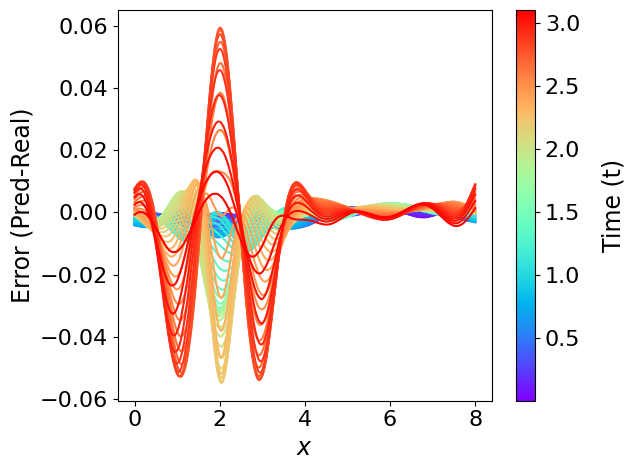}
    \caption{Time Evolution of Density from SPINN with potential obtained by solving Poisson equation through Fourier transform: density distributions are shown at fixed time frames shown by the color gradient with spatial coordinate in x-axis, and density in the y-axis.}
    \label{fig:Poiss}
\end{figure}

From Fig.~(\ref{fig:Poiss}), we observe that while this method allows the network to approximate solutions to the Schrödinger-Poisson system, the resulting errors are slightly higher than when the potential is predicted directly by the network as an auxiliary variable, as shown in Fig.~(\ref{Fig3}). Moreover, as discussed in the main text, solving for \(V\) via Fourier transforms introduces significant computational overhead, particularly for high-resolution grids. The increased memory and compute demands often force a reduction in grid size to avoid memory overflow, which in turn limits the achievable accuracy.

In contrast, predicting \(V\) alongside \(\psi\) within the network avoids these issues, leading to faster convergence, better scalability to high-resolution domains, and overall lower prediction errors.

\subsection{Different Activation Functions} \label{act_app}

\begin{figure*}
    \centering
    \begin{tabular}{cc}
     \includegraphics[width=0.5\linewidth]{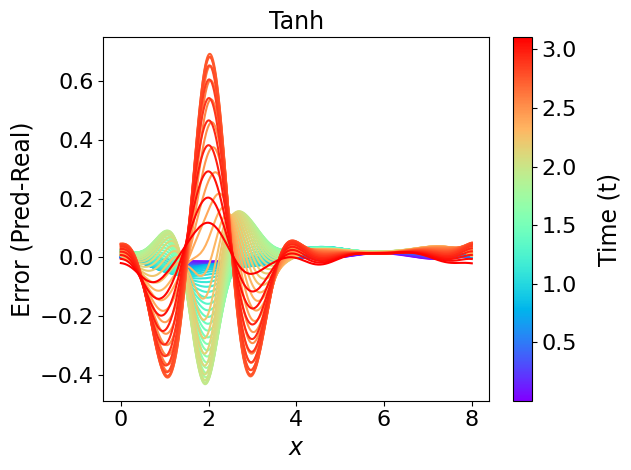}    &  \includegraphics[width=0.5\linewidth]{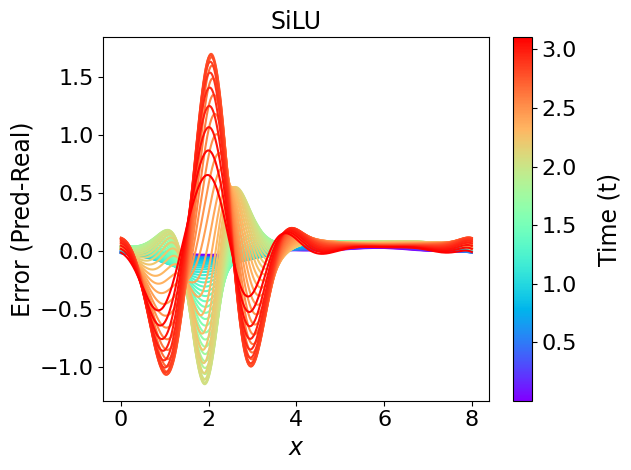}
    \end{tabular}
    \caption{Error comparison for a few activation functions: The plot depicts the error in density prediction on the y-axis vs the spatial coordinates at fixed time frames represented by color gradient, calculated w.r.t the numerical solution. }
    \label{fig:act}
\end{figure*}

In this subsection, we present results for SPINNs when different activation functions are used in place of the sine activation. Specifically, we train SPINNs using commonly employed activation functions in PINNs, such as \(\tanh\) and SiLU (the sigmoid-weighted linear unit).To ensure a fair comparison, We keep the network architecture, optimizer, and hyperparameters the same as in the 1D experiments in the main text, with the only change being the activation function.
As shown in Fig.~(\ref{fig:act}), the errors in the predicted density using \(\tanh\) and SiLU activations are noticeably higher compared to the case with sine activation, shown in Fig.~(\ref{Fig3}), highlighting the advantage of periodic functions like sine for oscillatory problems.

\end{document}